\documentclass[pra,twocolumn,showpacs,letterpaper,showpacs,superscriptaddress]{revtex4-1}

\usepackage{graphicx,amsmath,amssymb,amsfonts,latexsym,color,dcolumn,bm,epsfig,subfigure}
\usepackage[plainpages=false,hyperfootnotes=false,colorlinks=false]{hyperref}
\usepackage[normalem]{ulem}
\usepackage{dsfont}

\renewcommand{\imath}[0]{\mathrm{i}}
\newcommand{\abs}[1]{\left\vert#1\right\vert}

\renewcommand{\i}[2]{\int_{#1}^{#2}}

\begin{document}

\title{Spatial dispersion in atom-surface quantum friction} 

\author{D. Reiche}
\affiliation{Humboldt-Universit\"at zu Berlin, Institut f\"ur Physik, 
             AG Theoretische Optik \& Photonik, 12489 Berlin, Germany}
\affiliation{Max-Born-Institut, 12489 Berlin, Germany}

\author{D. A. R. Dalvit}
\affiliation{Theoretical Division, MS B213, Los Alamos National Laboratory, Los Alamos, NM 87545, USA}

\author{K. Busch}
\affiliation{Humboldt-Universit\"at zu Berlin, Institut f\"ur Physik, 
             AG Theoretische Optik \& Photonik, 12489 Berlin, Germany}
\affiliation{Max-Born-Institut, 12489 Berlin, Germany}

\author{F. Intravaia}
\affiliation{Max-Born-Institut, 12489 Berlin, Germany}

\newcommand{\mathbfh}[1]{\hat{\mathbf{#1}}}

\begin{abstract} 
We investigate the influence of spatial dispersion on atom-surface quantum friction.
We show that for atom-surface separations shorter than the carrier's mean free path within
the material, the frictional force can be several orders of magnitude larger than that predicted by local optics. 
In addition, when taking into account spatial dispersion effects, we show that the commonly used local thermal equilibrium approximation 
underestimates by approximately 95\% the drag force, obtained by employing the recently reported nonequilibrium fluctuation-dissipation relation for quantum friction. Unlike the treatment based on local optics, spatial dispersion in conjunction with corrections to local thermal equilibrium not only change the magnitude but also the distance scaling of quantum friction.
\end{abstract}

%\pacs{42.50.Ct, 12.20.-m, 78.20.Ci}
\maketitle

%%%%%%%%%%%%%

\section{Introduction} \label{sec:introduction}

Quantum fluctuations give rise to numerous fascinating physical effects, especially on
sub-micrometer scales. Some of these phenomena have been extensively studied and carefully
measured, thus demonstrating their relevance for both fundamental physics and future 
technologies \cite{Decca05,Intravaia13}. 
Recently, there has been a renewed interest in fluctuation-induced interactions  in nonequilibrium systems. 
A prominent example is quantum friction \cite{Pendry97,Volokitin07}, 
the quantum drag force between two uncharged, polarizable objects in relative motion.
A large part of the existing literature on quantum friction considers an atom (or some other microscopic object) moving in front of a flat surface, where the corresponding material is modeled using local optics, i.e. assuming an optical response described by a permittivity that only depends on frequency. \cite{Mahanty80,Schaich81,Tomassone97,Volokitin02,Dedkov02a,Kyasov02,Scheel09,Barton10b,Pieplow13,Intravaia14}. 
Within the assumption of local optics, several conceptual questions have been previously addressed, including the functional 
dependence of the frictional force on the atom's velocity \cite{Scheel09,Intravaia14,Intravaia15}, 
the impact of non-Markovian effects \cite{Intravaia16}, and even the relevance of nonequilibrium correlations \cite{Intravaia16a}. 

At short distances from the surface, however, a local description of the material becomes 
inadequate. Earlier works  \cite{Heinrichs75,Esquivel03,Sernelius05,Esquivel-Sirvent06} 
have already shown that using a \emph{nonlocal} description can lead to corrections to equilibrium 
dispersion forces \cite{Jackson75}. Nonlocality is to be understood in the sense that spatial dispersion is included in an optical response, i.e. the material's permittivity depends on both frequency and wavevector. Also, in the case of surface-surface quantum friction, different material models that include 
spatial dispersion have been used to describe the drag force \cite{Persson98,Volokitin03a,Volokitin07,Volokitin11,Despoja11,Despoja15}. 
The authors of these works have demonstrated that for short distances spatial nonlocality 
can lead to an enhancement of the force relative to the case of a local material model 
\cite{Persson98,Volokitin03a,Volokitin11,Despoja11}. 
These works, however, have resorted to the so-called local thermal equilibrium (LTE) 
approximation, where it is assumed that the subsystems in relative motion are at equilibrium 
with their immediate surroundings. 
Such a procedure allows to utilize results of equilibrium thermodynamics, like the fluctuation-dissipation theorem 
(FDT) \cite{Callen51}, but neglects the contribution due to nonequilibrium correlations \cite{Intravaia16a,Barton16}. 
In fact, recent work has shown that the LTE approximation is not well justified for  
atom-surface quantum friction and underestimates the magnitude of the drag force \cite{Intravaia16a}. 
Since the LTE approximation relies on a short correlation length of the fluctuations 
that mediate the interaction, one can expect that, when spatial dispersion is taken into 
account, the deviation from the LTE result is even larger than in the local optics treatment.

In this work, we study effects of spatial dispersion in atom-surface quantum friction and compare the results
obtained using the LTE approximation with those obtained using a full nonequilibrium approach. We show
that spatial dispersion enhances the failure of the LTE approximation, resulting in a 95\% deviation from the full nonequilibrium result compared 
to the 80\% deviation previously reported within local optics \cite{Intravaia16a}. 
In addition, we show that the inclusion of spatial nonlocality strongly 
affects the functional distance dependence of the frictional force in the low-velocity limit. In contrast to the local optics case, 
where both the LTE and the full nonequilibrium approach predict the same distance scaling law for the quantum frictional force, 
their distance behaviors are different in the presence of spatial dispersion.

\section{Atom-surface quantum friction}\label{secII}

Consider an atom driven by an external force and moving with non-relativistic 
velocity ($|\mathbf{v}|\ll c$) at constant height $z_{a}>0$ parallel to a conducting 
isotropic half-space. The atom is modeled as an electric dipole, described by the quantum operator $\mathbfh{d}(t)$.
% and, since the surface is invariant under rotations around the $z$-axis, we can, without loss of generality,
%assume that the motion is along the $x$-direction.
Due to the interaction of the atom with the surrounding quantum electromagnetic field a 
drag force will progressively balance the external drive until the system reaches a 
nonequilibrium steady state, where the motion continues with constant velocity. 
Dissipation in the material  gives rise to a nonzero memory 
time, such that in the nonequilibrium steady state we can ignore the transient acceleration process and assume 
that the atom has reached the trajectory $\mathbf{r}_{a}(t)=\mathbf{r}_{0}+v_xt\mathbf{x}$ 
\cite{Intravaia15,Intravaia16} (we assume that the motion is along the $x$-direction).
In an earlier work \cite{Intravaia16}, we have shown that the zero-temperature drag force felt by the atom in such a situation can be written as
\begin{align}\label{F}
F=-2&\i{0}{\infty}d\omega\int\frac{d^2\mathbf{p}}{(2\pi)^2}\nonumber\\
&\times p_x\text{tr}\left[\underline{S}_R(p_xv_x-\omega,v_x)\cdot\underline{G}_I^s(\mathbf{p},z_{a},\omega)\right],
\end{align}
where $p=|\mathbf{p}|=\sqrt{p_x^2+p_y^2}$ is the parallel component of the three-dimensional electromagnetic 
wave vector $\mathbf{k}=p_x\mathbf{x}+p_y\mathbf{y}+q\mathbf{z}$ 
\cite{Note1}. For symmetry reasons, 
the frictional force is only along the direction of the motion, i.e. $\mathbf{F}=F\mathbf{x}$. 
Quantum friction is determined by the velocity-dependent nonequilibrium power spectrum 
tensor of the dipole fluctuations, $\underline{S}(\omega,v_x)$, and by the Fourier transform 
(in time and $xy$-direction) of the electromagnetic surface Green tensor, 
$\underline{G}(\mathbf{p},z_{a},\omega)$. 
In Eq. \eqref{F} and in the remainder of the paper the subscript $R$ ($I$) denotes the 
real (imaginary) part of an expression and the superscript $s$ gives the symmetric part 
of a tensor \cite{Note1}. The Green tensor is given by the sum of a vacuum contribution $\underline{G}_0$ 
and a scattering contribution $\underline{g}$. Because of Lorentz invariance, only the 
latter contributes to the final result 
\cite{Dedkov03,Volokitin08,Pieplow13}. 
In all this work we focus on atom-surface distances within the surface's near-field region. In 
this case the part of the scattered Green tensor relevant to quantum friction \cite{Note1} is
\begin{align}
\label{NFGreen}
\underline{g}^{s}(\mathbf{p},z_{a},\omega)=\frac{pe^{-2z_{a}p}}{2\epsilon_0}r(\omega,p)\left[\frac{p_x^2}{p^2}\mathbf{x}\mathbf{x}+\frac{p_y^2}{p^2}\mathbf{y}\mathbf{y}+\mathbf{z}\mathbf{z}\right]~,
\end{align}
where $\epsilon_0$ is the vacuum permittivity. The description of the material properties 
enters via the transverse magnetic reflection coefficient, $r(\omega,p)$, which in general 
depends on both the frequency and, for symmetry reasons, the modulus of the wavevector 
$\mathbf{p}$. In a spatially local description of the material and in the near-field limit, 
the dependence on the wave vector disappears and the reflection coefficient is only a 
function of frequency 
\cite{Pieplow13,Intravaia16}.

In order to calculate the nonequilibrium power spectrum, we model the dipole's internal 
dynamics as a harmonic oscillator \cite{Intravaia16} 
\begin{equation}
\partial^{2}_{t}\mathbfh{d}(t)+\omega_{a}^{2}\mathbfh{d}(t)=\omega_{a}^{2}\underline{\alpha}_{0}\cdot \mathbfh{E}(\mathbf{r}_a(t),t), 
\label{eqnMotionOsc}
\end{equation}
where $\omega_{a}$ is the oscillator's frequency corresponding to the atom's characteristic dipolar resonance frequency
\cite{Note2}, $\mathbfh{E}$ is the electric field, 
and $\underline{\alpha}_{0}$ is the static polarizability tensor, assumed to be symmetric 
for simplicity (it is proportional to a projector parallel to the direction of the dipole 
moment). We suppose that the oscillator has no intrinsic dissipation and that all the 
dissipative dynamics arises from the coupling to the electromagnetic field.
The harmonic oscillator model allows for an analytical expression of $\underline{S}(\omega,v_x)$ 
given by  \cite{Intravaia16a}
\begin{align}
\label{SnonLTE}
\underline{S}(\omega,v_x)&=\frac{\hbar}{\pi}\left[\theta(\omega)\underline{\alpha}_I(\omega,v_x)+\underline{J}(\omega,v_x)\right],
\end{align}
where $\theta(\omega)$ is the Heaviside-theta function. In contrast to the LTE approach which relies on the equilibrium FDT, this nonequilibrium FDT \eqref{SnonLTE} contains the extra term 
\begin{align}\label{J}
\begin{split}
\underline{J}(\omega,v_x)=&\int \frac{d^2\mathbf{p}}{(2\pi)^2}\left[\theta(\omega+p_xv_x)-\theta(\omega)\right]\\
\times&\underline{\alpha}(\omega,v_x)\cdot\underline{G}_I(\mathbf{p},z_{a},\omega+p_xv_x)\cdot\underline{\alpha}^*(\omega,v_x).
\end{split}
\end{align}
In the previous equations
\begin{equation}
\underline{\alpha}(\omega,v_x)=
\frac{\omega_{a}^{2}}{ \omega_{a}^2  -\Delta(\omega,v_x)-\omega^{2}-\imath \omega \gamma(\omega,v_x)}\,\underline{\alpha}_{0} 
\label{alpha}
\end{equation}
is the velocity-dependent atomic polarizability, where $\gamma(\omega,v_x)$ 
and $\Delta(\omega,v_x)$ denote, respectively, the velocity-dependent radiative damping 
and frequency shift \cite{Klatt16}
\begin{subequations}
\begin{align}
\frac{\Delta (\omega,v_{x})}{\omega_{a}^{2}}=\int \frac{d^{2}\mathbf{p}}{(2\pi)^{2}}\mathrm{Tr}\left[\underline{\alpha}_{0}\cdot\underline{G}_{R}(\mathbf{p},z_{a}, \omega+p_xv_x)\right]&,\\
\label{gammarate}
\frac{\omega\gamma(\omega,v_{x})}{\omega_{a}^{2}}=\int \frac{d^{2}\mathbf{p}}{(2\pi)^{2}}\mathrm{Tr}\left[\underline{\alpha}_{0}\cdot\underline{G}_{I}(\mathbf{p},z_{a},\omega+p_xv_x)\right] &.
\end{align}
\end{subequations} 
According to Eq. \eqref{SnonLTE}, the frictional force in Eq. \eqref{F} decomposes into 
two contributions, 
\begin{equation}
F=F^{\rm LTE}+ F^{J}.
\end{equation}
The first, $F^{\rm LTE}$, is what one would 
have obtained by applying the LTE approximation, while the second, $F^{J}$, is the 
correction entirely due to the nonequilibrium dynamics of the system. 

Previous works 
\cite{Volokitin08,Volokitin15a,Silveirinha14a,Intravaia16,Intravaia16a} 
have shown that the quantum frictional process is characterized by a non-resonant and 
a resonant contribution, both being a function of the atomic velocity and the atom's 
separation from the surface. The resonant part occurs for sufficiently high velocities 
which bring the atomic transition frequency  within the range of the surface plasmon-polariton 
resonances that exist at the vacuum/material interface. Here, we consider only the 
non-resonant part of the frictional force which takes place at lower velocities
and is more likely to play a central role 
in typical experimental setups.
In Appendix \ref{Low-velocity} we show that the main contribution to the force comes 
from the frequency range $0<\omega\lesssim v_{x}/z_{a}$ (see also Refs. 
\cite{Intravaia16,Intravaia16a}). Therefore, at sufficiently low velocities 
\cite{Note3} the drag force is determined by the low-frequency behavior of the material's electromagnetic 
response. Under the assumption that the material is Ohmic for these low frequencies 
(we will see below that this applies to our nonlocal model), the low-velocity approximation 
of the LTE and the nonequilibrium contributions to the friction can be written as (see Appendix \ref{Low-velocity})
\begin{subequations}
\begin{align}
F^{\rm LTE}&\approx- 2\hbar\frac{v_{x}^{3}}{\pi} \frac{\Phi_{0}\Phi_{2}}{3}\frac{\mathcal{D}_{0}(z_{a})\mathcal{D}_{2}(z_{a})}{\left[1-\Delta(0,0)/\omega_{a}^{2}\right]^{2}},
\label{generalAsympLTE}\\
F^{J}&\approx- 2\hbar\frac{v_{x}^{3}}{\pi}\Phi_{1}^{2}\frac{\mathcal{D}^{2}_{1}(z_{a})}{\left[1-\Delta(0,0)/\omega_{a}^{2}\right]^{2}}.
\end{align}\label{generalAsymp}
\end{subequations}
This shows that at low velocities the zero-temperature frictional force grows as the third power of the atom's velocity 
\cite{Volokitin02,Dedkov02a,Kyasov02,Pieplow13,Intravaia14}.
In the above expressions, we have introduced the abbreviations
\begin{equation}
\Phi_{n}=\binom{2n}{n}\frac{\frac{2n+1}{2(n+1)} \alpha_{xx} + \frac{1}{2(n+1)} \alpha_{yy} + \alpha_{zz}}{2^{2n+3}\pi\epsilon_0}~,
\label{Geometry}
\end{equation}
associated with the dipole's direction in space, and 
\begin{equation}
\mathcal{D}_{n}(z_{a})=\int_{0}^{\infty}dp\,
p^{2 (n+1)}e^{-2z_{a}p}
r'_{I}(0,p),
\label{DistanceDep}
\end{equation}
which depends on the properties of the surface (the prime indicates the first derivative with 
respect to the frequency). The functions $\mathcal{D}_{n}(z_{a})$ are the [$2n$]-th derivative with respect to $z_{a}$ of the 
low-frequency behavior of the electromagnetic density of states near the vacuum/material interface.
In particular, $\mathcal{D}_{0}(z_{a})$ is related to the atomic decay rate induced by the 
interaction with the radiation (radiative damping). Eqs. \eqref{generalAsymp} show that, 
under the assumption of Ohmic dissipation, the LTE and the nonequilibrium 
correction have the same functional dependency on the velocity, while their behavior as a function of the distance
can be distinct. In the local optics approximation, however,  we have that 
$\mathcal{D}_{0}(z_{a})\propto z_a^{-3}$, 
$\mathcal{D}_{1}(z_{a})\propto z_a^{-5}$, and
$\mathcal{D}_{2}(z_{a})\propto z_a^{-7}$
(see Eq. \eqref{locallimitD} in Appendix \ref{nonlocalimpedance}). In this case $F^{\rm LTE}$ and $F^{J}$ have the same $z_a^{-10}$ distance dependency, as was already shown in  \cite{Intravaia16a}.

%%%%%%%%%%%%% Material Model %%%%%%%%%%%%%%%%%

\section{The spatially dispersive material model for the metallic bulk}\label{sec:model}

The previous results allow for a quantitative evaluation of the impact of spatial dispersion
on quantum friction. 
At this point, we would like to recall that spatial dispersion becomes physically relevant for materials in which the
free-carriers can move over distances which are much larger than the interatomic separation. This 
extreme mobility of charged particles is also related to collective phenomena, such as plasmon oscillations in metals \cite{Pines56}, dynamical screening \cite{Feibelman82,Liebsch87} and quantum many-body phenomena \cite{Pines66}. In a macroscopic continuum description of the material, spatial dispersion leads to a nonlocal relation between the displacement and the electric fields, leading to a permittivity that depends on the wavevector of the radiation \cite{Jackson75}.
In this paper we focus on a metallic surface and describe its properties using the so-called 
semi-classical infinite barrier (SCIB) model 
\cite{Lindhard54,Ford84}. 
In this model, electrons are treated as a Fermi fluid whose dynamics is governed by 
the Boltzmann equation. At interfaces, electrons are assumed to be specularly reflected 
by an infinite potential barrier \cite{Kliewer67,Kliewer68}. 
Although more sophisticated models are available (see, for example, Refs. \cite{Intravaia15a,Feibelman82}), 
the SCIB model takes into account important phenomena, such as Landau damping 
\cite{Landau46,Van-Kampen55}, which are absent in simpler nonlocal models (e.g. the hydrodynamic model) 
\cite{Feibelman82,Intravaia15}. 
Landau damping occurs when the frequency and the wavevector of the radiation fulfill the 
condition $\omega\approx\mathbf{k}\cdot\mathbf{v}_{p}$, i.e., when the quasi-particle's velocity 
$\mathbf{v}_{p}$ becomes comparable to the phase velocity $\mathbf{v}_{ph}$ of the radiation, 
$v_{p}\sim v_{ph}=\omega/k$ ($k=|\mathbf{k}|$). 
%\new{In the modern language of open quantum system, this process is equivalent to the ``trace over''
%of the continuum of degrees of freedom corresponding to the quasi-particle velocitiies in the phase 
%space \cite{Van-Kampen55}.}
Since quantum friction is very sensitive to any form of dissipation present in the system 
\cite{Intravaia14,Intravaia16}, 
this intrinsic damping due to the exchange of energy between the electronic wave 
function and the radiation \cite{Pines66} will play an important role in our analysis.

Within the SCIB model, the reflection coefficient takes the form \cite{Ford84}
\begin{align}
r(\omega,p)=\frac{1-Z(\omega,p)/Z_0(\omega,p)}{1+Z(\omega,p)/Z_0(\omega,p)},
\label{R-Impedance}
\end{align}
where $Z(\omega,p)$ is the transverse magnetic surface impedance and $Z_0(\omega,p)$ is the 
corresponding vacuum value. In the non-retarded limit (formally equivalent to the limit for 
$c\to \infty$) we have 
\cite{Ford84,Haakh12}
\begin{align}\label{surfaceimpedance}
\frac{Z(\omega,p)}{Z_{0}(\omega,p)}\approx\frac{2}{\pi}\i{0}{\infty}dq~\frac{1}{k^{2}}\frac{p}{\epsilon_l(\omega,\mathbf{k})},
\end{align}
such that the reflection coefficient only depends on the longitudinal part of the bulk dielectric function, $\epsilon_l(\omega,\mathbf{k})$. For the latter we use the semi-classical limit of Lindhard's quantum dielectric function 
\cite{Lindhard54,Dressel02,Pines66}

\begin{align}
\epsilon_l(\omega,\mathbf{k})=1+\frac{\omega_{p}^{2}}{\omega+\imath\Gamma}\frac{3 u^2f_l(u)}{\omega+\imath\Gamma f_l(u)},
\label{epsilonl}
\end{align}
where $\Gamma$ is the metal's dissipation rate, $\omega_{p}$ is the plasma frequency, the function $f_l(u)$ reads
\begin{align}
 f_l(u)=1-\frac{u}{2}\ln\left[\frac{u+1}{u-1}\right] ,
\label{fl}
\end{align}
and $u=(\omega+\imath\Gamma)/(v_Fk)$ with $v_{F}$ the Fermi velocity.
Equation (\ref{epsilonl}) is obtained within linear response theory \cite{Lindhard54,Dressel02,Pines66} by assuming a thermal equilibrium Fermi-Dirac carrier distribution. Furthermore, the expression for the permittivity is valid for wavevectors much smaller than the Fermi wavevector $k_{F}=m_{e}v_{F}/\hbar$ 
($m_{e}$ the effective electron mass) or, equivalently, when the wavelength 
of the radiation is much larger than the de Broglie wavelength $\lambda_{B}$ of the electron at the 
Fermi surface \cite{Mermin70,Ashcroft76,Chapuis08}. Deviations from a Fermi-Dirac distribution have to be considered for strong interactions occurring at time scales shorter than the carrier equilibration time $\omega_{p}^{-1}\lesssim \tau_{\rm Eq} \lesssim \Gamma^{-1} $ (usually shorter than 1 ps in metals) \cite{Pines66,Fann92}.
In addition, corrections to the semi-classical approach are expected for
atom-surface separations $z_{a}\ll\lambda_{B}/\pi$, which for metals corresponds to half the Bohr radius, i.e., 
few tenths of an \AA ngstr\"om. As explained in section \ref{secII}, quantum friction is a weak low frequency phenomenon and, therefore, by considering distances $z_{a}>1\text{ \AA}$ our approach is well within the range of validity of such a description.

Depending on the value of $u$, different mechanisms dominate the optical response of 
the metal. In the limit $\abs{u}\to\infty$ we recover the local Drude model
$\epsilon_{l}(\omega,k)\to\epsilon_{\rm D}(\omega)=1-\omega_{p}^{2}[\omega(\omega+\imath\Gamma)]^{-1}$.
In this case the main contributions to the atom-surface interaction stem from wavelengths 
$\lambda\sim 1/k$ much larger than the electron's mean free path, $\lambda\gg\ell=v_F/\Gamma$. 
For typical physical parameters $\ell$ ranges from a few tens up to a few hundreds of nanometers. 
Since $v_{F}/c$ is of the order of the fine-structure constant, we then obtain $\ell\approx 50$ nm for a gold bulk with
$\Gamma\sim 30$ meV. The local (Drude) regime corresponds to a situation where the electrons' dynamics 
averages over a multi-scattering scenario and their ballistic motion is negligible. 
In this limit the phase velocity $v_{ph}$ becomes larger than the Fermi velocity $v_F$, 
inhibiting the interaction responsible for Landau damping \cite{Pines66} (see also below). 
On the other hand, for $\abs{u}\to 0$, the wave resolves the ballistic motion of 
the electron ($\lambda\ll \ell$), leading to a distinct spatially dispersive response to the 
electromagnetic field \cite{Jackson75}.
Scattering becomes less relevant and, since the phase velocity of the radiation is smaller or 
equal than the Fermi velocity, Landau damping takes over as the dominant damping mechanism.
Mathematically, this phenomenon is represented by the imaginary part of the function $f_l(u)$ in the limit
$\abs{u}\to 0$ due to the logarithm appearing in Eq. \eqref{fl} (see also Eq. \eqref{flexpansion} in Appendix \ref{nonlocalimpedance}).

The same physical mechanisms determine the behavior of the surface impedance. For 
$p \ell\ll\abs{\omega/\Gamma+\imath}$ we have $Z(\omega,p)/Z_0(\omega,p)\approx 1/\epsilon_{\rm D}(\omega)$ 
which leads to the usual local limit for the reflection coefficient 
\cite{Intravaia14}. 
In the limit $p\ell\gg\abs{\omega/\Gamma+\imath}$ spatial dispersion is relevant and in Appendix 
\ref{nonlocalimpedance} we show that we can write
\begin{align}\label{ZNonlocal}
\frac{Z(\omega,p)}{Z_0(\omega,p)}\approx\frac{p\lambda_{\rm TF}}{\sqrt{1+p^{2}\lambda_{\rm TF}^{2}}}-\imath\frac{\frac{\omega}{\omega_{p}}Q\left(p\lambda_{\rm TF},\frac{\pi^2-4}{2\pi p\ell}\right)}{p \lambda_{\rm TF}(1+p^{2}\lambda_{\rm TF}^{2})},
\end{align}
where we have defined the function
\begin{equation}
Q(a,b)=\frac{a^{2} \left(a^2+1\right)}{\sqrt{3}}\int_{0}^{1}dx~\frac{1+b x}{ \sqrt{1-x^{2}}}\frac{x^3}{(a^{2}+x^2)^{2}}.
\label{QFunction}
\end{equation}
Note that the function $Q(a,b)$ is real and nonzero also for $b=0$, which corresponds to 
the limits $\ell \to \infty$ or $\Gamma\to 0$. Indeed, due to the Landau damping and despite a vanishing collision rate $\Gamma$, 
Eq. \eqref{ZNonlocal} still has a nonzero imaginary part at low frequencies, which implies a dissipative reflection coefficient.
In Eq. \eqref{ZNonlocal} $\lambda_{\rm TF}=v_{F}/(\sqrt{3}\omega_{p})$ is the Thomas-Fermi screening length, which is on 
the order of few \AA ngstr\"oms for typical metals \cite{Ashcroft76} (see also Appendix \ref{nonlocalimpedance}) and characterizes the electrostatic screening of charges in the Fermi fluid (it can be considered as the analog of the Debye 
length at zero temperature \cite{Ashcroft76,Manfredi05}). In our system $\lambda_{\rm TF}$ is also related to
spatial distribution of the  electron density near the surface \cite{Heinrichs73a}.
Within the nonlocal region, the values for which $p\lambda_{\rm TF}<1$ correspond to electromagnetic 
waves that propagate with a phase velocity $v_{ph}>v_{F}$ in the metal. Since at zero temperature 
no particles exist with velocity larger than the Fermi velocity, in this region both, Landau 
damping and impurity scattering, are concurrent but not fully effective dissipative processes. 
Conversely, for $p\lambda_{\rm TF}>1$ dissipation is dominated by the interaction 
between the electrons and the electromagnetic waves. From Eqs. \eqref{R-Impedance} and 
\eqref{ZNonlocal} and for $v_{x}\ll v_{F}$ we have that the imaginary part of the reflection 
coefficient is
\begin{equation}
r_{I}(\omega,p)\approx \frac{2\frac{\omega}{\omega_{p}}Q\left(p\lambda_{\rm TF},\frac{\pi^2-4}{2\pi p\ell}\right)}{p \lambda_{\rm TF}\left(\sqrt{1+p^{2}\lambda_{\rm TF}^{2}}+p\lambda_{\rm TF}\right)^{2}}\equiv2\omega\epsilon_{0}\rho(p),
\label{resistivity}
\end{equation}
showing that the material has an Ohmic behavior and that, formally, the resistivity $\rho(p)$ depends on the wave vector when spatial dispersion is relevant.
The function $\rho(p)$ grows as $\abs{p \lambda_{\rm TF} \ln (p \lambda_{\rm TF})}$ for small 
wave vectors, features a maximum around $p\sim 1/(5 \lambda_{\rm TF})$, and then decreases as 
a power law for large $p\lambda_{\rm TF}$ (see Appendix \ref{nonlocalimpedance}). Importantly, 
at its maximum $\rho(p)$ can be more than an order of magnitude larger than the typical resistivity of a metal in the local optics description, $\rho=\Gamma/(\epsilon_{0}\omega_{p}^{2})$. We show in the next Section that the previous characteristics deeply impact the distance dependency of quantum friction between an atom and a spatially dispersive metal.

\section{Effects of spatial dispersion on quantum friction}

\begin{figure}[t]
\center
\includegraphics[width=8.8cm]{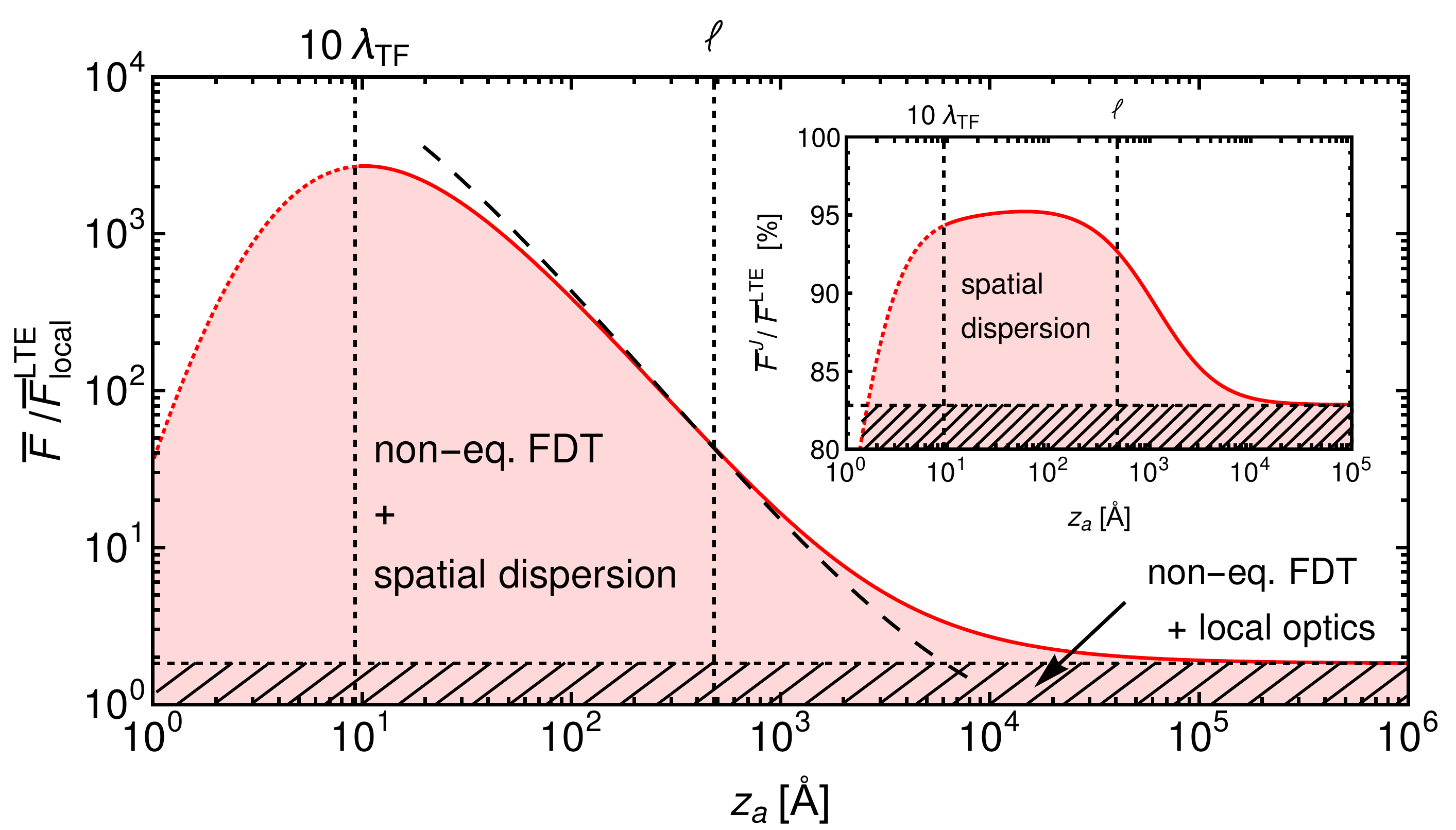}
\caption{Quantum frictional force acting on an atom moving at constant velocity and parallel to a metallic surface described by the SCIB model. The low-velocity limit of the force (Eqs. \eqref{generalAsymp}) $\bar{F}$ is plotted as a function of the atom-surface separation $z_a$. In all plots, the parameters $\omega_{p}=9$ eV, $\Gamma=30$ meV, and $v_{F}/c=1/137$ are fixed to these same values.
In order to emphasize the role of spatial dispersion and the nonequilibrium physics, the force $\bar{F}$ is normalized to its expression obtained using local optics and the LTE approximation, $\bar{F}^{\rm LTE}_{\rm local}$. 
In the low-velocity regime the normalized force does not dependend on the velocity (see Eqs. \eqref{generalAsymp} and \eqref{LocalLimit}).  
For atom-surface separations much larger than the electron's mean free path $z_a \gg \ell=v_{F}/\gamma$, the force approaches a value which is almost twice that of $\bar{F}^{\rm LTE}_{\rm local}$, recovering the result reported in Ref. \cite{Intravaia16a}.
For $z_a < \ell$, spatial dispersion and the non-LTE correction result in a substantial increase of the force, with a maximum enhancement at $z_a\sim 10\lambda_{\rm TF}$, where $\lambda_{\rm TF}=v_{F}/(\sqrt{3}\omega_p)$ is the Thomas-Fermi screening length. The curve is dotted for distances less than 10 \AA, where our description might not be reliable (see text). The black dashed line shows the total asymptotic behavior for distances $\lambda_{\rm TF}\ll z_{a}<\ell$  given by the sum of the expressions in Eqs. \eqref{nonLocalLimit}. The inset shows the correction exclusively due to the nonequilibrium physics, i.e. $\bar{F}^{J}/\bar{F}^{\rm LTE}$  with spatial dispersion taken into account in both forces of the numerator and the denominator. The curve shows a larger contribution of the non-LTE correction in the nonlocal case than in the local limit. It also indicates that, for the parameters used here, nonlocality is the main source of the force enhancement observed for $z_a < \ell$.
}
\label{figQF}
\end{figure}

Combining Eq. \eqref{resistivity} with Eqs. (\ref{generalAsymp}-\ref{DistanceDep}), we are able 
to compute the low-velocity quantum frictional force including spatial dispersion in the material's optical response. The results are presented in Fig. \ref{figQF}, where the force is normalized with 
respect to $\bar{F}^{\rm LTE}_{\rm local}$, which is the force obtained by the simultaneous use of local optics and LTE approximations (see Eq. \eqref{LocalLimit} below). The normalization is chosen in order to highlight the impact of the non-LTE corrections and of the nonlocal material  properties. We also perform an average over the dipole's spatial directions (denoted by the bars above the forces) and define $\alpha_{0}=\mathrm{Tr}[\underline{\alpha}_{0}]/3$, which coincides with the expression of the static isotropic atomic polarizability. 
According to Sec. \ref{sec:model}, we can distinguish between three physically different 
regions of atom-surface separations. 

For distances $z_{a}\gg\ell$, local optics is a valid description of the metal. 
From Eqs. \eqref{generalAsymp} we recover the results for quantum friction obtained in Ref. 
\cite{Intravaia16a} 
(see also the end of the Appendix \ref{nonlocalimpedance})
\begin{align}
\bar{F}_{\rm local}^{\rm LTE}
\approx- \hbar\frac{189}{2\pi^{3}} \left( \frac{\alpha_{0}}{\epsilon_0}\frac{\Gamma}{\omega_{p}^{2}}\right)^{2}
\frac{ v_{x}^{3}}{(2 z_{a})^{10}},\quad
\frac{\bar{F}_{\rm local}^{J}}{\bar{F}_{\rm local}^{\rm LTE}}
\approx \frac{29}{35}~.
\label{LocalLimit}
\end{align}
Here, for simplicity, we neglected the contribution originating from the frequency shift, 
which for distances $z_{a}\gg (\alpha_0/\epsilon_0)^{1/3}$ (few \AA ngstr\"oms 
for typical atoms) gives only a subleading contribution to the force that arises from the term $[1-\Delta (0,0)/\omega _{a}^{2}]$ in Eqs. (\ref {generalAsymp}).

When spatial dispersion is relevant, i.e. for separations smaller than the electron's mean free path, 
we can identify two distinct distance regimes. Starting with $\lambda_{\rm TF}\ll z_{a}< \ell$, 
we obtain
\begin{subequations}
\begin{gather}
\frac{\bar{F}^{\rm LTE}}{\bar{F}_{\rm local}^{\rm LTE}}
\approx \frac{\omega_{p}^{2}}{\Gamma^{2}}
\frac{\left[\ln\left(\frac{B_0 z_a}{\lambda_{\rm TF}}\right)+\frac{C_0 z_a}{\ell}\right]\left[\ln\left(\frac{B_2 z_a}{\lambda_{\rm TF}}\right)+\frac{C_2 z_a}{\ell}\right]}{\frac{1} {7}\left(\frac{2z_a}{\lambda_{\rm TF}}\right)^{2}}~,
\label{nonLocalLimitLTE}
\\
\frac{\bar{F}^{J}}{\bar{F}_{\rm local}^{\rm LTE}}
\approx \frac{145}{7} \frac{\omega_{p}^{2}}{\Gamma^{2}}
\left[\frac{\ln\left(\frac{B_1 z_a}{\lambda_{\rm TF}}\right)+\frac{C_1 z_a}{\ell}}{\frac{2z_a}{\lambda_{\rm TF}}}\right]^{2},
\end{gather}
\label{nonLocalLimit}
\end{subequations}
where $B_{n}$ and $C_{n}$ are the following numerical constants: $B_0\approx0.69$, 
$B_1\approx 0.44$, $B_2\approx0.32$, $C_0\approx0.98$, $C_1\approx0.59$, and $C_2\approx0.42$
(see also Appendix \ref{nonlocalimpedance}).
We note that spatial nonlocality induces a non-algebraic change in the distance dependence of the 
force and, in contrast to the local optics case, the distance scalings of $\bar{F}^{\rm LTE}$ and $\bar{F}^{J}$ are different. As it was expected from the considerations regarding the system's nonequilibrium 
dynamics (see Sec. \ref{sec:introduction}), the contribution of the term $\bar{F}^{J}$ to 
the full frictional force is larger than in the local case, inducing a correction that reaches 
about 95 \% rather than 80 \% of the LTE contribution (see inset of Fig. \ref{figQF}). 
Importantly, the full nonequilibrium force in the nonlocal case is larger than the corresponding local counterpart calculated for values of the damping rate $\Gamma=30$ meV and of the plasma frequency $\omega_{p}=9$ eV. 
For both the LTE and the nonequilibrium contribution, nonlocality leads to an increase in the force that scales with $\omega_{p}^{2}/\Gamma^{2}$ (see Eqs. \eqref{nonLocalLimit}). Therefore, it is particularly relevant for very clean materials.
The largest enhancement of roughly three orders of magnitude is reported for a distance 
$z_{a}\sim 10 \lambda_{\rm TF}$ (of the order of 1 nm for typical metals; see Fig \ref{figQF}). 
This value is effectively an additional intrinsic length scale of the system which derives 
from the combination of geometry and material properties. It is related to the value of $z_{a}$ 
for which the functions $p^{2 (n+1)}e^{-2z_{a}p}$ and $r'_{I}(0,p)$ (see Eq. \eqref{resistivity}) 
appearing in the integral defining $\mathcal{D}_{n}(z_{a})$ have the maximum overlap. 
Physically speaking, $p^{2 (n+1)}e^{-2z_{a}p}$ selects as a function of the distance $z_{a}$
the parallel component of the wavevectors participating in the dissipative process in the
material described by $r'_{I}(0,p)$. For each $n$ this occurs at $z_{a}\approx 5(n+1)\lambda_{\rm TF}$. 
Another interesting point to note is the influence of spatial dispersion on the metal's resistivity. 
The magnitude of the frictional force at $z_{a}\sim 10 \lambda_{\rm TF}$ is equivalent to that obtained via local optics but with a much larger dissipation rate of $\Gamma\sim 1 $ eV. This corresponds to a $\sim$30 times higher resistivity, showing the relevance of spatial dispersion on quantum friction. This behavior can be understood with more 
detail by looking at the wavevector-dependent resistivity implicitly defined in 
Eq. \eqref{resistivity}: As described above, for values of $p\ell\gg 1$, $\rho(p)$ can reach values much 
larger than those of the local optics description.

%%%%%%%

%%%%%%%

For distances $z_{a}\ll \lambda_{\rm TF}$, the functional behavior of the frictional force changes once
again. Although our model allows for a full mathematical 
characterization of the interaction in this distance range, atomic scale effects, dynamical screening and electron spill-out become relevant at such short separations (shorter than an {\AA}ngstr\"om for usual metals), and the continuum description of the materials is no longer reliable. For recent discussions on these topics see \cite{Zhu16b,Teperik13,Raza15}. Nevertheless, it is worth mentioning that for 
distances $z_{a}\lesssim 10\lambda_{\rm TF}$ the frictional force (denoted by the dotted line in Fig. \ref{figQF}) is still stronger, but increases more slowly than its local counterpart (see Fig. \ref{figQF}). In this case the 
overlap in Eq. \eqref{DistanceDep} for $\mathcal{D}_{n}(z_{a})$ selects wavevectors for
which the dissipative process described by the resistivity in Eq. \eqref{resistivity} 
becomes less effective.

Finally, we comment on the difference between the behavior of the frictional force 
for microscopic systems discussed so far, where dissipation is induced by the interaction with the electromagnetic field, and for systems where the source of dissipation is 
internal, like for instance in metallic nanoparticles. 
As pointed out in previous work \cite{Intravaia16}, the LTE approximation usually 
provides the leading contribution to the frictional force for the latter case, for which internal 
dissipation is much stronger than radiative damping. Within our treatment, such 
systems can be described by a polarizability like in Eq. \eqref{alpha}, but with a 
vanishing frequency-shift, $\Delta(\omega,v_x)=0$ and a constant damping rate 
$\gamma(\omega,v_x)\equiv \gamma$. At low velocities the force is given by the relation
\begin{equation}
F^{\rm LTE}\approx- 2\hbar\frac{v_{x}^{3}}{\pi} \frac{\Phi_{2}}{3} \frac{\gamma}{\omega_{a}^{2}}\mathcal{D}_{2}(z_{a})~.
\label{intrinsic}
\end{equation}
When compared with Eq. \eqref{generalAsympLTE}, the previous expression shows a difference 
in the functional dependence on the distance. In the local case this corresponds to a change 
in the exponent of the power law from $z_{a}^{-10}$ for radiative damping to $z_{a}^{-7}$ 
for intrinsic damping \cite{Intravaia14} (see Eq. \eqref{locallimitD}). In the spatially dispersive
case using the SCIB model, however, a more qualitative modification of the functional 
behavior occurs. For distances $\lambda_{\rm TF}\ll z_{a}< \ell$ one has
\begin{equation}
\frac{\bar{F}^{\rm LTE}}{\bar{F}_{\rm local}^{\rm LTE}}
\approx \frac{7}{\sqrt{3}}\frac{\omega_{p}}{\Gamma}\frac{\ln\left[\frac{B_{2}z_a}{\lambda_{\rm TF}}\right]+\frac{C_{2}z_a}{\ell}}{\frac{2 z_{a}}{\lambda_{\rm TF}}}
\label{intrinsicEnh}
\end{equation}
which shows that, in addition to a 
change in the power law exponent, the system with intrinsic dissipation features a single 
logarithm instead of the product of two logarithms as obtained in Eq. \eqref{nonLocalLimitLTE} 
(see also Eqs. \eqref{nonlocallimitD} and \eqref{locallimitD}). For such separations Eq. \eqref{intrinsic} also reveals 
an enhancement of the interaction due to spatial dispersion. The strengthening of the nonlocal 
frictional force with respect to its local counterpart has a maximum around $z_{a}\approx 15\lambda_{\rm TF}$. 
However, in the case of intrinsic dissipation, the force enhancement is less significant because
$\bar{F}^{\rm LTE}/\bar{F}^{\rm LTE}_{\rm local}$ in Eq. \eqref{intrinsicEnh} is proportional to $\omega_{p}/\Gamma$ and not to 
its square (see Eq. \eqref{nonLocalLimitLTE}). 

%%%%%%%%%%%%%%%

\section{Conclusions}

In the present work we investigated the impact of spatial dispersion on atom-surface quantum 
friction for non-relativistic velocities. Our description goes beyond the widely used local 
thermal equilibrium approximation and does not rely on the usual equilibrium fluctuation-dissipation theorem, 
but rather on an extension of it for nonequilibrium steady states \cite{Intravaia16a}. The analysis focuses on 
the behavior of the frictional force for small velocities, which 
are more likely to be achieved in experimental setups.
We show that for distances shorter than the electron's mean free path $\ell$, spatial dispersion 
and the system's nonequilibrium processes have a large impact on quantum friction enhancing 
the interaction with respect to the LTE value. 
The closer the atom gets to the surface, the less important the collision-induced damping
becomes and the more the Landau damping takes over as source of dissipation (see Sec. \ref{sec:model}). 
A maximum enhancement of three order of magnitude is attained for distances that are of the 
order of ten times the Thomas-Fermi screening length $\lambda_{\rm TF}$. 
Our results also show that in the nonlocal system the failure of the LTE approximation is 
more significant than in the local system, and underestimates the force by about 95 \% (the 
nonequilibrium processes are responsible for half of the total frictional force; see the inset of Fig. \ref{figQF}).
The inclusion of spatial dispersion does not alter the functional dependence of the interaction 
on the atomic velocity, which is proportional to $v_{x}^{3}$, but it deeply modifies its behavior as a function of the distance. 
Physically, this difference can be understood by recalling that the velocity dependence is related to the low frequency 
behavior of the electromagnetic density of states \cite{Intravaia14,Intravaia16}, Ohmic for both the spatial dispersive and local materials (a sub-ohmic or a super-ohmic material will also affect the functional dependency on the velocity). Instead, the behavior as a function of the atom-surface separation is more related to the detail of the medium's optical response and to the different length scales associated with the physical processes occurring in the material.
For atom-surface separations $\lambda_{\rm TF}\ll z_{a}< \ell$, quantum friction is no longer described by a simple power 
law but, due to Landau damping, it acquires a more complex structure involving a logarithmic 
contribution and the combination of length scales $\ell$ and $\lambda_{\rm TF}$ (see 
Eqs.\eqref{nonLocalLimit} and \eqref{intrinsicEnh}). In addition, unlike the local case, the contribution to the force resulting from the LTE approximation and its correction have a different distance dependence, showing again the relevance of the interplay of nonequilibrium effects and spatial dispersion for quantum frictional processes. 

Quantum friction is a very weak effect and experimental investigations are therefore challenging \cite{Intravaia16}. Relatively simple time-of-flight experiments, where atoms are sent parallel to a surface and decelerations or stopping distances are measured, are possible but they may not provide the required sensitivity. Consequently, rather sophisticated atom-interferometric techniques would be better suited \cite{Cronin09, Impens13}. In view of the desired atom-surface separations, such experimental designs come with their own challenges. For instance, at least one arm of the interferometer must be aligned parallel to the surface at comparatively short distances of some tens of nanometers to the surface. The frictional force will produce a different phase accumulation in this arm with respect to a second arm placed at much larger separations from the surface. The resulting phase shift, encoding the information on the drag force, will appear in the interference pattern. In order to exploit the enhancement effects associated with spatial nonlocality, the atom-surface separation should be of the order of or shorter than the electron mean free path $\ell$ in the material composing the surface (large $\ell$ values correspond to clean materials). 
While this clearly is challenging, it is not entirely out of reach.
Inspecting Eqs. \eqref{LocalLimit} and \eqref{nonLocalLimit}, we note that in the nonlocal region the frictional force scales in absolute value as $\omega_{p}^{-2}$. According to our results, preferable characteristics of the surface material are therefore a reasonably small dissipation rate as well as plasma frequency. 
The latter conditions point, for example, to doped semiconductors like ZnO:Ga for which $\Gamma\sim 50$ meV and $\omega_{p} \sim 1$ eV have already been measured \cite{Sadofev13}.
The dielectric response of highly-doped semiconductors is, however, more involved than the simple model used here and the corresponding behavior of quantum friction will be investigated in detail in future work.

\section{Acknowledgments} 
DR thanks the CNLS for the hospitality as well as for financial support and the DAAD for funding 
through the PROMOS program. We also acknowledge support by the LANL LDRD program, and by 
the Deutsche Forschungsgemeinschaft (DFG) through project B10 within the Collaborative Research 
Center (CRC) 951 Hybrid Inorganic/Organic Systems for Opto-Electronics (HIOS). FI further 
acknowledges financial support from the European Union Marie Curie People program through the 
Career Integration Grant No. PCIG14- GA-2013-631571 and from the DFG through the DIP program 
(Grant No. SCHM 1049/7-1).

\appendix

\section{Low-velocity expansion}
\label{Low-velocity}

In this appendix we  provide the main steps of the procedure that allows to obtain the 
low-velocity asymptotic expressions given in Eq. \eqref{generalAsymp}. First of all, it is 
convenient to define the tensor
\begin{equation}
\underline{\mathcal{G}}(\abs{p_{x}},z_{a},\omega)=\int_{-\infty}^{\infty} \frac{dp_{y}}{2\pi}
\underline{G}(\mathbf{p},z_{a},\omega)~,
\end{equation}
which has the same symmetry properties as $\underline{G}(\mathbf{p},z_{a},\omega)$ with 
respect to the variable $\omega$, namely an even real part and an odd imaginary part under 
the change $\omega\to -\omega$. 
Inserting Eq. \eqref{SnonLTE} in Eq. \eqref{F} we can define
\begin{subequations}
\begin{multline}
F^{\rm LTE}=- 4\hbar\int_{0}^{\infty}\frac{dp_{x}}{2\pi}
 p_x \int_{0}^{p_{x}v_{x}}\frac{d\omega}{2\pi}\\
\times \text{Tr}\left[\underline{\alpha}_{I}(p_xv_x-\omega,v_x)\cdot\underline{\mathcal{G}}_{I}(\abs{p_{x}},z_{a},\omega)\right]~,
\end{multline}
\begin{multline}
F^{J}
 =- 2\hbar\int_{-\infty}^{\infty}\frac{dp_{x}}{2\pi}
 p_x \int_{-\infty}^{\infty}\frac{d\omega}{2\pi}\\
 \times\text{Tr}\left[\underline{J}(p_xv_x-\omega,v_x)\cdot\underline{\mathcal{G}}_{I}(\abs{p_{x}},z_{a},\omega)\right]~.
\label{FJ}
\end{multline}
\end{subequations}
Further manipulations of the previous expressions are possible. First, using the parity 
properties of $\underline{\mathcal{G}}(\abs{p_{x}},z_{a},\omega)$ and treating the cases $\omega>0$ 
and $\omega<0$ separately, one can show that the expression for $\underline{J}(\omega,v_x)$ in Eq. \eqref{J} 
can be rewritten as (see also the Supplementary Material of Ref. \cite{Intravaia16})
\begin{multline}
\underline{J}(\omega,v_x)=\int_{\frac{\abs{\omega}}{v_{x}}}^{\infty} \frac{dp_{x}}{2\pi}\\
\times\underline{\alpha}(\omega,v_x)\cdot
\underline{\mathcal{G}}_{I}(\abs{p_{x}},z_{a},p_xv_x-\abs{\omega})\cdot\underline{\alpha}^*(\omega,v_x)~.
\end{multline}
\begin{widetext}
The integration in Eq. \eqref{FJ} can be simplified by rearranging the integration domain 
as follows
\begin{equation}
\int_{-\infty}^{\infty}\frac{d\omega}{2\pi}\int_{\frac{\abs{p_xv_x-\omega}}{v_{x}}}^{\infty} \frac{d\tilde{p}_{x}}{2\pi}
(\ldots) =\int_{0}^{\infty} \frac{d\tilde{p}_{x}}{2\pi}\int_{p_{x}v_{x}}^{(p_{x}+\tilde{p}_{x})v_{x}}\frac{d\omega}{2\pi} (\ldots) 
+\int_{0}^{\infty} \frac{d\tilde{p}_{x}}{2\pi}\int_{(p_{x}-\tilde{p}_{x})v_{x}}^{p_{x}v_{x}}\frac{d\omega}{2\pi} (\ldots)~.
\end{equation}
Combining all the previous expressions, using the parity properties of $\underline{\mathcal{G}}(\abs{p_{x}},z_{a},\omega)$ and the definition of $\gamma(\omega, v_{x})$ in Eq. \eqref{gammarate} we have
\begin{subequations}
\begin{align}
F^{\rm LTE}
 =- 4\hbar\int_{0}^{\infty}\frac{dp_{x}}{2\pi}
 p_x \int_{-\infty}^{\infty} \frac{d\tilde{p}_{x}}{2\pi}
 \int_{0}^{p_{x}v_{x}}&\frac{d\omega}{2\pi}
 \abs{A(p_xv_x-\omega,v_x)}^{2}\nonumber\\
&\times \mathrm{Tr}\left[\underline{\alpha}_{0}\cdot\underline{\mathcal{G}}_{I}(\abs{\tilde{p}_{x}},z_{a},[p_x+\tilde{p}_x]v_x-\omega)\right]\text{Tr}\left[\underline{\alpha}_{0}\cdot\underline{\mathcal{G}}_{I}(\abs{p_{x}},z_{a},\omega)\right]~,\\
F^{J}
 =- 2\hbar\int_{-\infty}^{\infty}\frac{dp_{x}}{2\pi}
 p_x\int_{-\infty}^{\infty} \frac{d\tilde{p}_{x}}{2\pi}\int_{p_{x}v_{x}}^{[p_{x}+\tilde{p}_{x}]v_{x}}&\frac{d\omega}{2\pi}
 \abs{A(p_xv_x-\omega,v_x)}^{2}\nonumber\\
&\times\text{Tr}\left[
\underline{\alpha}_{0}\cdot
\underline{\mathcal{G}}_{I}(\abs{\tilde{p}_{x}},z_{a},[p_x+\tilde{p}_x]v_x-\omega)\cdot\underline{\alpha}_{0}\cdot\underline{\mathcal{G}}_{I}(\abs{p_{x}},z_{a},\omega)\right]~,
\end{align}
\end{subequations}
\end{widetext}
where we also rewrote Eq. \eqref{alpha} as $\underline{\alpha}(\omega,v_x)=A(\omega,v_x) \underline{\alpha}_{0}$.
Due to the exponential function in the Green tensor in Eq. \eqref{NFGreen}, the dominant wavevectors contributing to the above integrals are $p_{x}\lesssim 1/z_{a}$. 
The previous expressions show that quantum friction is dominated by frequencies $\omega \lesssim v_{x}/z_{a}$.
Under the assumption that for these frequencies a Taylor expansion in $\omega$ of the function 
$\underline{\mathcal{G}}(\abs{p_{x}},z_{a},\omega)$ is possible and that
$\underline{\mathcal{G}}(\abs{p_{x}},z_{a},\omega)\approx\underline{\mathcal{G}}_{R}(\abs{p_{x}},z_{a},0)+\imath \omega \underline{\mathcal{G}}'_{I}(\abs{p_{x}},z_{a},0)$ describes the relevant physics, we have
\begin{widetext}
\begin{subequations}
\begin{align}
F^{\rm LTE}
&=- 2\frac{v_{x}^{3}}{\pi}\hbar \abs{A(0,0)}^{2}\frac{1}{3}\int_{0}^{\infty}\frac{dp_{x}}{2\pi}
p_x^{4} \text{Tr}\left[\underline{\alpha}_{0}\cdot\underline{\mathcal{G}}'_{I}(\abs{p_{x}},z_{a},0)\right]\int_{0}^{\infty} \frac{d\tilde{p}_{x}}{2\pi}
 \mathrm{Tr}\left[\underline{\alpha}_{0}\cdot\underline{\mathcal{G}}'_{I}(\abs{\tilde{p}_{x}},z_{a},0)\right]~,\\
F^{J}
&=- 2\frac{v_{x}^{3}}{\pi}\hbar\abs{A(0,0)}^{2}\int_{0}^{\infty}\frac{dp_{x}}{2\pi}
 p_x^{2} \text{Tr}\left[\underline{\alpha}_{0}\cdot\underline{\mathcal{G}}'_{I}(\abs{p_{x}},z_{a},0)\right]
 \int_{0}^{\infty} \frac{d\tilde{p}_{x}}{2\pi}\tilde{p}_x^{2}
 \mathrm{Tr}\left[\underline{\alpha}_{0}\cdot\underline{\mathcal{G}}'_{I}(\abs{\tilde{p}_{x}},z_{a},0)\right]~,
\end{align}
\end{subequations}
\end{widetext}
where we also used that $\underline{\alpha}_{0}=2\mathbf{d}\mathbf{d}/\hbar \omega_{a}$. The 
previous expressions are quite similar and they involve products of the integrals
\begin{align}
I_{n}(z_{a})=\int_{0}^{\infty}\frac{dp_{x}}{2\pi}
p_x^{2 n} \text{Tr}\left[\underline{\alpha}_{0}\cdot\underline{\mathcal{G}}'_{I}(\abs{p_{x}},z_{a},0)\right]~,
\end{align}
where $n=0,1,2$. Using the expression for the Green tensor given in Eq. \eqref{NFGreen}, after 
going to polar coordinates one can show that the previous function can be written as 
$I_{n}(z_{a})=\Phi_{n}\mathcal{D}_{n}(z_{a})$, where $\Phi_{n}$ 
and $\mathcal{D}_{n}(z_{a})$ are given by
\begin{equation}
\Phi_{n}=\binom{2n}{n}\frac{\frac{2n+1}{2(n+1)} \alpha_{xx} + \frac{1}{2(n+1)} \alpha_{yy} + \alpha_{zz}}{2^{2n+3}\pi\epsilon_0}~,
\end{equation}
and
\begin{equation}
\mathcal{D}_{n}(z_{a})=\int_{0}^{\infty}dp\,
p^{2 (n+1)}e^{-2z_{a}p}
r'_{I}(0,p).
\end{equation}

%%%%%%%%%

\section{The nonlocal impedance}
\label{nonlocalimpedance}

In this appendix we discuss the equations reported in Sec. \ref{sec:model}.
For this purpose it is convenient to define the following quantities:
$\ell=v_{F}/\gamma$, $\lambda_{\rm TF}=v_{F}/(\omega_{p}\sqrt{3})$ and $\kappa_{F}=(\omega+\imath \gamma)/v_{F}=\omega/v_{F}+\imath/\ell$. 
Using these definitions we rewrite the dielectric function in Eq. \eqref{epsilonl} as 
\begin{align}
  \epsilon_l(\kappa_{F},u)
  &=\left[1+\frac{u^2}{\kappa_{F}^{2}\lambda_{\rm TF}^{2}}\right]\left[1+\frac{u^2}{\kappa_{F}^{2}\lambda_{\rm TF}^{2}+ u^2}\frac{f_l\left(u\right)-1}{1+\imath \frac{f_l(u)}{\kappa_{F}\ell-\imath}}\right]~,
 \end{align}
where $u=\kappa_{F}/k$.
In the first factor we recognize the Thomas-Fermi dielectric function 
$\epsilon_{\rm TF}(k)=1+(k^{2}\lambda_{\rm TF}^{2})^{-1}$ describing charge screening in the metal 
\cite{Ashcroft76}. 
The additional factor is the correction introduced by the semi-classical Lindhard dielectric
function 
\cite{Ford84}. 
As explained in the main text (see the beginning of Sec. \ref{sec:model}) the nonlocal
region is characterized by $\abs{u}\ll 1$ (low frequencies and/or large wavevectors) and therefore
\begin{equation}
f_{l}(u)=1+\frac{\imath \pi  u}{2}-u^{2}+\mathcal{O}(u^{4}).
\label{flexpansion}
\end{equation}
The accuracy of the previous approximation also explains why the asymptotic behaviors presented in Eqs. \eqref{nonLocalLimit} provide a good description also for $z_{a}\lesssim \ell$.
In this regime the correction to the Thomas-Fermi model is small \cite{Barton79,Halevi95,Kittel96} and with the changes of variable $q \to u=x\kappa_{F}/p$, 
we can write Eq. \eqref{surfaceimpedance} as
\begin{align}
\label{surfaceimpedanceapp}
\frac{Z(p,\omega)}{Z_{0}(p,\omega)}&\approx\frac{2}{\pi}\int_{0}^{1}dx~\frac{1}{ \sqrt{1-x^{2}}}\frac{1}{\epsilon_l\left(\kappa_{F},x\frac{\kappa _F}{p}\right)}~.
\end{align}
In the nonlocal region $\kappa_{F}/p$ is small and expanding the integrand in this 
variable leads to
\begin{align}
\frac{1}{\epsilon_l\left(\kappa_{F},x\frac{\kappa _F}{p}\right)}&\approx
\frac{1}{1+\frac{x^2}{p^{2}\lambda_{\rm TF}^{2}}}\nonumber\\
&-\imath \frac{\frac{\pi}{2}\frac{\omega}{\omega_{p}\sqrt{3}}p \lambda_{\rm TF} x^3}{(p^{2}\lambda_{\rm TF}^{2}+x^2)^{2}}\left[1+x\frac{\pi^2-4}{2 \pi p\ell}\right]~.
\end{align}
In the above expression we have neglected a contribution $\propto \omega^{2}$ since for 
quantum friction we are interested in the low-frequency (Ohmic) region. The above 
expression allows for an analytical evaluation of Eq. \eqref{surfaceimpedanceapp}. 
The relevant integrals are
\begin{equation}
\frac{2}{\pi}\int_{0}^{1}dx~\frac{1}{ \sqrt{1-x^{2}}}\frac{1}{1+\frac{x^2}{a^{2}}}=\frac{a}{\sqrt{1+a^{2}}}
\end{equation}
and the integral given in Eq. \eqref{QFunction}, defining the function $Q(a,b)$. Although 
it results in a mathematically more involved function, this last integral can also be evaluated 
analytically and gives
\begin{align}
Q(a,b)&=\frac{a^{2}}{2\sqrt{3}}\bigg{[}\frac{a^2+2}{\sqrt{a^2+1}}\ln\left(\frac{\sqrt{a^{2}+1}+1}{a}\right)-1\nonumber\\
&\hspace{2cm}+\left(a^2+1-a\frac{a^2+\frac{3}{2}}{ \sqrt{a^2+1}}\right) \pi b\bigg{]}\nonumber\\
&\approx\frac{2}{\sqrt{3}}
\begin{cases}
-\frac{a^2}{2}  \left[\ln \left(\frac{a}{2}\right)+\frac{1-\pi b}{2}\right]& a\ll 1\\
\frac{1}{3}-\frac{1}{5a^2}+\frac{3}{32}\left(1-\frac{2}{3a^2}\right) \pi  b& a\gg 1
\end{cases}
~.
\end{align}
These results lead to Eq. \eqref{resistivity} and allow the evaluation of the asymptotic expressions 
of the function $\mathcal{D}_{n}(z_{a})$. In the nonlocal region we distinguish the cases 
$n=0$ and $n=1,2$:
\begin{subequations}
\begin{equation}
\mathcal{D}_{0}(z_{a})\approx
\begin{cases}
\frac{4\sqrt{3}\lambda_{\rm TF}}{(2z_a)^4}\frac{\ln\left(\frac{B_{0}z_a}{\lambda_{\rm TF}}\right)+\frac{C_{0}z_a}{\ell}}{\omega_{p}}& z_{a}\gg \lambda_{\rm TF}\\
-\frac{1}{3\sqrt{3}\lambda_{\rm TF}^{3}}\frac{\ln\left(\frac{\tilde{B}_{0} z_{a}}{\lambda_{\rm TF}}\right)-\frac{\frac{\pi z_{a}}{\sqrt{2}}}{\lambda_{\rm TF}}}{\omega_{p}}& z_{a}\ll \lambda_{\rm TF}
\end{cases}~,
\end{equation}
\begin{equation}
\mathcal{D}_{n}(z_{a})\approx
\begin{cases}
\frac{2\lambda_{\rm TF}}{\sqrt{3}}\frac{(2n+3)! }{(2 z_{a})^{2 n+4}}\frac{\ln\left(\frac{B_{n}z_a}{\lambda_{\rm TF}}\right)+\frac{C_{n}z_a}{\ell}}{\omega_{p}}& z_{a}\gg \lambda_{\rm TF}\\
\frac{(2n-1)!}{3\sqrt{3}\lambda_{\rm TF}^{3}(2 z_{a})^{2n}} \frac{1}{\omega_{p}}& z_{a}\ll \lambda_{\rm TF}
\end{cases}~,
\end{equation}
\label{nonlocallimitD}
\end{subequations}
which lead to the the expression reported in Eq. \eqref{nonLocalLimit}. 
We defined the constants $B_{n}=4\exp(\Gamma_{\rm Eu}-f_n)$, with $f_n=7/3,167/60,433/140$, 
and $C_{n}=(\pi^2-4)/h_n$, with $h_{n}=6,10,14$, where $\Gamma_{\rm Eu}\approx 0.58$ is the Euler 
constant. These are the numerical values that are reported after Eq. \eqref{nonLocalLimit} in the
main text. In addition we defined $\tilde{B}_{0}=\sqrt{2}\exp\left[\Gamma_{\rm Eu}\right]\approx 2.52$.

In the local region ($z_a \gg \ell$), we obtain
\begin{equation}
\mathcal{D}_{n}(z_{a})\stackrel{z_{a}\gg \ell}{\approx} 2\frac{\Gamma}{\omega_{p}^{2}}
\begin{cases}
\frac{2!}{(2 z_{a})^{3}} & n=0\\
\frac{4!}{(2 z_{a})^{5}} & n=1\\
\frac{6!}{(2 z_{a})^{7}} & n=2
\end{cases}~,
\label{locallimitD}
\end{equation}
which lead to the expressions in Eq. \eqref{LocalLimit}.

%%%%%%%%%%%%%%%%%%%

%\bibliographystyle{/Users/nabu/Documents/Mydocs/Lavoro/bibliography/bibstyle/prstytitle}
%\bibliography{/Users/nabu/Documents/Mydocs/Lavoro/bibliography/biblio}

\end{document}